# Role of Layer Thickness and Field-Effect Mobility on Photoresponsivity of Indium Selenide (InSe) Based Phototransistors

Milinda Wasala[1,2*], Prasanna Patil[1*], Sujoy Ghosh[1], Lincoln Weber[1], Sidong Lei[3], Saikat Talapatra[1]

[1]Department of Physics, Southern Illinois University, Carbondale, IL 62901, USA.
[2]Science Department, Great Basin College, Elko, NV 89801, USA.
[3]Department of Physics and Astronomy, Georgia State University, Atlanta, GA 30303, USA.

*Equally contributing authors
#corresponding author; Email: saikat@siu.edu

## Abstract

Understanding and optimizing the properties of photoactive two-dimensional (2D) Van der Waals solids are crucial for developing optoelectronics applications. Here we present a detailed investigation of layer dependent photoconductive behavior of InSe based field-effect transistors (FETs). InSe based FETs with five different channel thickness (t, 20 nm < t < 100 nm) were investigated with a continuous laser source of λ = 658 nm (1.88 eV) over a wide range of illumination power of 22.8 nW < P < 1.29 µW. All the devices studied, showed signatures of photogating, however, our investigations suggest that the photoresponsivities are strongly dependent on the thickness of the conductive channel. A correlation between the field-effect mobility ($\mu_{FE}$) values (as a function of channel thickness, t) and photoresponsivity (R) indicates that in general R increases with increasing $\mu_{FE}$ (decreasing t) and vice versa. The maximum responsivity of ~ 7.84 A/W and ~ 0.59 A/W was obtained for the device with t = 20 nm and t = 100 nm respectively. These values could substantially increase under the application of a gate voltage. The structure–property correlation-based studies presented here indicate the possibility of tuning the optical properties of InSe based photo-FETs for a variety of applications related to photodetector and/or active layers in solar cells.

**Introduction**

With the advent of layered Van der Waals solids [1-3] and the envisioned applications that are being expected [4-13] has led to a large variety of fundamental as well as applied scientific investigations. Among these investigations, many have shown that intrinsic properties of 2D layered materials are strongly dependent on the number of layers [8,14-16]. For example, on one hand 2D layered semiconductors such as Molybdenum Disulfide ($MoS_2$), is a direct band material in monolayer form and is an indirect band material in its multilayer form and on the other hand materials such as Indium Selenide (InSe) is a direct bandgap material in its multilayer form and an indirect band material in monolayer form. Such effect on the intrinsic properties due to simple addition of layers also influences functional properties of devices such as field-effect transistors (FETs) [17,18]. For example, in FETs fabricated using $MoS_2$ [17,19] and InSe [20,21], the field-effect mobility ($\mu_{FE}$) shows as inverse relationship with material thickness. Similarly, these studies also indicate that on/off ratios decrease drastically with increasing layer thickness. Additionally, the electronic interaction of 2D conductive channels with underlying substrate(for example silicon dioxide ($SiO_2$)) indicate that fundamental phenomena such as charge impurity screening depends on the layer numbers in $MoS_2$ transistors [18].

Although the intrinsic properties of 2D Van der Waals solids vary with layer thickness, some core properties such as the knowledge of the nature of the bandgap in 2D semiconducting layers are crucial for developing function specific applications. For example, although $MoS_2$ is a direct band-gap material in its monolayer form, its applicability in photoconductive driven phenomenon could be limited due to the difficulty in fabricating/obtaining consistently monolayer material as well as the limitation in the photo absorption process due to its single atom thick layer. Recent studies

[8,22,23] have also shown that multilayer direct band gap materials can serve better in applications where higher photo absorption is needed for device applications leading to photodetectors, photo-FETs/switch. In that respect, it is now well established that, multilayer InSe layers with number of layers > ~6 (or thickness t > ~5 nm) will fall under materials with direct band gap [15,24]. However, to harness the specific functional aspect of photodetection, a better understanding of how the photoconductive properties behave with increasing number of layers is needed. In this article, results of a detailed investigations on photoconductive properties of five InSe based FET devices are presented. The thickness of the conductive channel of these devices ranges between 20nm to 100nm (~24 – 120 layers). Our findings indicate that upon light illumination some key fundamental phenomenon for example shift in the threshold voltage, photogating effect etc. occurred in all the devices tested. However, we found that the thinner samples, which typically showed higher field-effect mobility, had higher photoresponsivities. The values of the photoresponsivity obtained for the FET device with channel thickness ~20 nm is more than two orders of magnitude higher than the device with channel thicknesses ~ 80-100 nm. These investigations shed light on the critical issue of identifying and optimizing one of the most significant material parameters e.g., the structure-property correlation of layered 2D materials, essential for the development of future technological applications.

**Materials and Method**

The Bulk $\gamma$-InSe crystals were synthesized by thermal treatment of a nonstoichiometric mix of Indium and Selenium. Details of crystal synthesis and structural characterization of bulk InSe were reported elsewhere [25]. Figure 1(a) shows the layer stacking of $\gamma$-InSe crystal. Se-atoms in the first layer aligned with the In-atoms in the second layer and Se-atoms in the second layer aligned with the In-atoms in the third layer. The layer height of InSe crystal was measured as ~0.84 nm and the interlayer distance between the two adjacent layers of InSe were measured as ~0.29 nm [8]. To perform the structural and opto-electronic characteristics of the few layers thick $\gamma$-InSe flakes, bulk InSe crystals were mechanically exfoliated on to p+ doped Si wafers with at least ~ 300 nm thermally grown SiO$_2$ layer. The inset of the Figure 1(b) shows a scanning electron microscopy (SEM) image of one of the mechanically exfoliated $\gamma$-InSe flake. Energy dispersive X-ray spectrum (EDX) of the exfoliated $\gamma$-InSe flake showing clear peaks for In & Se is presented in the Figure 1(b).

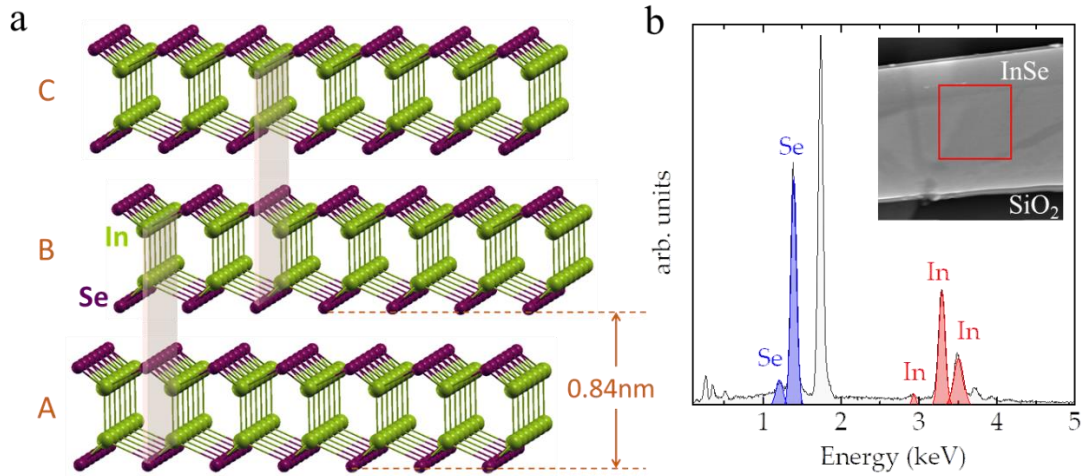

**Figure 1.** (a) Layer staking of $\gamma$-InSe (green spheres- In atoms, violet spheres- Se atoms). (b) SEM image of mechanically exfoliated InSe flake on Si/SiO$_2$ substrate (inset) and EDX spectra of the area (red square) of InSe flake shown in the inset. Dominant In and Se peak showed in red and blue color respectively.

In order to perform electrical and opto-electronic characterizations, chromium (Cr ~ 40 nm) and gold (Au ~ 160 nm) electrodes were deposited on the selected flakes using a shadow mask via thermal evaporation. Optical image of electrodes deposited on InSe device 2 (D2) is shown in Figure 2(a), where the optical active area of the device was calculated as ~ $2.1 \times 10^{-9}\ m^2$. The height of the InSe flakes were measured using the contact mode atomic force microscopy (AFM). Height of the D2 was found to be ~ 34 $nm$ (which corresponds to ~ 40 InSe layers) as shown in Figure 2(b). AFM measurements also show flat topography along the exfoliated top surface, which suggests the unique thickness of the sample. The samples were then mounted on ceramic chip holders with 3 terminal configuration for electrical/opto-electronic measurement where Cr/Au electrodes act as the source (S), drain (D) and Si/SiO$_2$ wafer act as the back-gate (G) (Figure 2(c)). A continuous laser illumination (Coherent Inc., CUBE) with a wavelength, λ = 658 nm, which corresponds to the energy, E = 1.88 eV and tunable laser illumination intensity was used to illuminate the devices. A laser with a spot size of diameter ~ 2.8 mm was used in the experiments.

**Results and Discussions**

We have fabricated and tested the photocurrent transport properties of five FET devices with different layer thickness t = 20 nm (D1), 34 nm (D2), 75 nm (D5), 80 nm (D6) and 100 nm (D7). Detailed analysis and mechanisms of electronic transport in these devices are mentioned elsewhere [20]. Room temperature transfer characteristics curve is shown in Figure 2(d). Within the applied gate voltage range (-60 V < V$_{GS}$ < 60 V) these InSe FET shows *n*-type behavior. Blue curve refers to the logarithmic scale (axis on the left) of the I$_{DS}$ and the red solid line refers to the linear scale (axis on the right) of the I$_{DS}$. The transfer characteristics curve was used to extract few figures of merit related to the FET. For example, with the application of V$_{DS}$ = 0.1 V, the field-effect mobility (µ$_{FE}$) of the device was found to be ~ 13.48 cm$^2$V$^{-1}$s$^{-1}$, calculated using the equation;

$$\mu_{FE} = \frac{L\ g_m}{W C_{ox} V_{DS}}$$

Where $C_{ox}$ is 3.4 x 10$^8$ F for SiO$_2$ thickness of 1000 nm (for D2). *L* & *W* are the channel dimensions and transconductance $g_m$ is the slope of the linear region of I$_{DS}$ – V$_{GS}$ curve. The ON/OFF ratio for this device was ~$10^4$. In Figure 2(e), we present the $I_{DS}$ vs $V_{DS}$ curves obtained at room temperature. Within the measurement range of $V_{DS}$, drain current shows almost linear behavior even with application of different $V_{GS}$.

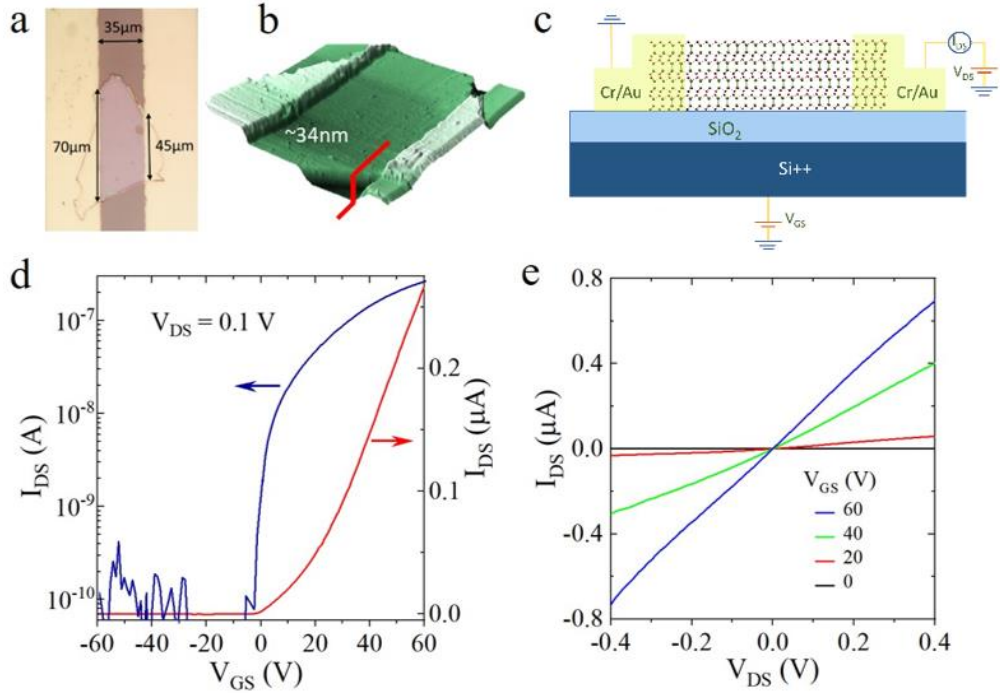

**Figure 2.** InSe FET Device D2 (a) Optical image of InSe FET D2 device. (b) 3D constructed image of the AFM height profile of the device. (c) Schematic diagram of the FET with back-gated configuration. (d) Room temperature transfer characteristics curve of the device D2. Blue curve refers to the logarithmic value of the drain current (Y-axis on the left) and red curve refers to the linear values of the drain current (Y axis on the right). (e) Room temperature output characteristics curve for different $V_{GS}$ values.

In Figure 3, results of the room temperature photocurrent measurements performed on this device is presented. The photocurrent measurements were performed using a continuous-wave laser source of wavelength $\lambda = 658$nm (E = 1.88eV) with a spot size of diameter ~ 2.8mm. The calculated area of the laser spot ~ $6.16 \times 10^{-6}\ m^2$, which was ~ 3 orders of magnitude greater than the area of the device. Larger illumination compared to the device is essential in order to equally illuminate the source and drain contact thereby minimizing the photo-thermoelectric effects [26]. Figure 3(a) shows the InSe FET (D2) device's room temperature transfer curves obtained with a source-drain voltage at 100 mV under dark conditions as well as under laser illumination (with increasing powers). For the data shown in Figure 3(a), the laser powers were varied from 0.07mW to 4mW, which equates to an effective laser illumination power ($P_{eff}$) from 29nW to 1.29µW. With the increment of laser illumination powers, the ON state drain currents ($I_{DS}$) were increased. For example, the ON state current increases from 0.266 µA (dark current at $V_{GS}$= +60 V) up to 0.718 µA at an illumination intensity of $P_{eff}$ ~ 1.29 µW. We also noticed that the threshold voltage ($V_{TH}$), extrapolated from the linear portion of the transfer curves, shifts towards the negative gate voltage region with increase in the laser intensity. The variation in the threshold voltage shift ($\Delta V_{TH}$), can be calculated using $\Delta V_{TH} = V_{TH} - V_{TH,Dark}$, as a function of laser illumination, is shown in the inset of Figure 3(a). The threshold voltage shift increases with increasing $P_{eff}$ and can be explained based on the presence of surface/interface traps states. It is generally believed that some of the photo-generated charge carriers are trapped or accumulated inside relatively long-lived trap states or recombination centers and do not participates directly in the photoconduction process [27]. These trapped charge carriers act as a source of additional gating effect and thus increases the majority carrier concentration inside the channel. This phenomenon, known as photogating effect, is often responsible for the shift of the threshold voltage upon illumination as previously reported in devices based on $MoS_2$ [28,29], $In_2Se_3$ [23] as well as in $CuIn_7Se_{11}$ [22]. We believe that the shift in the threshold voltages observed in InSe FET devices is due to the trap-mediated photogating effect.

In Figure 3(b), the variation of photocurrent ($I_{photo} = I_{illu} - I_{dark}$) as a function of $P_{eff}$ and under the application of different back-gate voltages (-60 V < $V_{GS}$ < 60 V) is presented. We found that, at a particular value of $V_{GS}$, $I_{photo}$ increases with increasing laser powers and for a fixed value of $P_{eff}$,

$I_{photo}$ increases with increasing applied $V_{GS}$. The nature of the variation of photocurrent with laser intensity under a varying gate voltage can provide insights into the mechanism of photocurrent generation mechanisms in low-dimensional materials. We found that all the curves obtained in Figure 3(b) can be fitted with a simple power law variation of the type, $I_{photo} \propto P_{eff}^{\gamma}$. For each applied gate voltage, the power exponent ($\gamma$) obtained from the fit is shown in Figure 3(c). We noticed that at lower (negative) gate voltages, $\gamma \sim 1$. As we increase the gate voltage, $\gamma \rightarrow 0$. For example, at $V_{GS} = -10$ V, $\gamma \sim 1$ and for $V_{GS} > 0$, the power exponent decreases to $\gamma \sim 0.3$. Such fractional power law dependence is very common in photodetectors [30] based on several 2D materials such as multilayer $In_2Se_3$ [23], $CuIn_7Se_{11}$ [22] etc.. Variation of the photocurrent upon light illumination can be understood if we consider that in case of pure photoconduction, $I_{photo}$ should be linearly dependent on the incident light illumination. However, presence of surface traps as well as their dynamics upon light illumination can lead to fractional values of the power exponents. For example, as reported earlier, in the case of $In_2Se_3$ based 2D photodetectors, traps could result from natural surface oxide present in the basal plane layers of the material due to the vacancies [31]. Similarly, surface oxides can also cause hole trapping as seen in ZnO and GaN nanowire photodetectors [32,33]. In some cases, these trapped charges produce a field of their own and act as an extra gate potential leading to photogating behavior. This is manifested through the fractional values of the power exponents. Therefore, the variation of the values of $\gamma$ from 1 at low gate voltages to $\gamma < 1$ at high voltages, observed in the InSe devices investigated here, is due to the manipulation of the fundamental mechanism of photocurrent generation in these materials under the application of gate voltage. We believe that the mechanism of photocurrent generation deviates from purely photoconductive at lower gate voltages (manifested through values of $\gamma \sim 1$) to trap mediated photogating at higher gate voltages (with $\gamma < 1$).

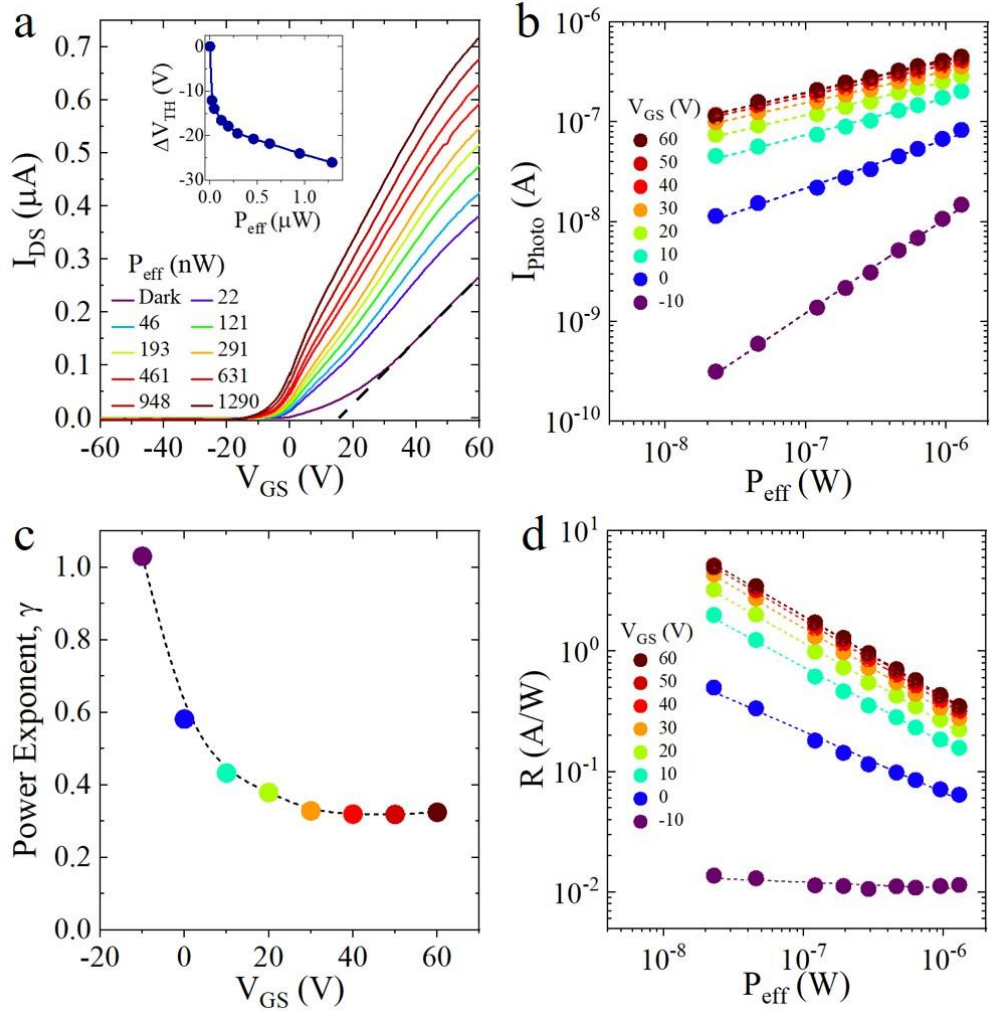

**Figure 3.** (a) Room temperature transfer curves of D2 at $V_{DS}$ = 0.1V for different laser illumination power of a 658 nm laser. The effective laser powers increase from 22nW to 1.29µW. (Inset) Threshold voltage shift for increasing effective laser powers. (b) Shows the variation of $I_{photo}$ as a function of effective laser intensity (log‑log scale) and at different gate voltages ($V_{GS}$ from −60 V to +60 V). (c) The variation of the power exponent (γ) as a function of gate voltage is presented. (d) Responsivity as a function of laser power and at different gate voltages is presented.

We have calculated the photoresponsivity (*R*) of D2. Photoresponsivity (*R*) is the ratio between photocurrent generated and total incident optical power on the device. This gives the measure of the achievable electrical signal under certain light illumination powers and can be extracted using equation:

$$R = \frac{I_{photo}}{P_{in}}$$

Where, $P_{in}$ is the incident light illumination power onto the device. For our measurements, spot size of light source used to calculate $R$ is larger than the size of the channel area of the device. Therefore instead of $P_{in}$, effective illumination power, $P_{eff}$, ($P_{eff} = P_{in} \cdot A_{device} / A_{spot}$) was used to estimate the $R$. Here, $A_{device}$, $A_{spot}$ are the device area (effective channel area) and the total area of the light source or the spot size, respectively. For this device, we found that room temperature responsivity at lowest illumination power with $V_{GS} = 0$ V and $V_{DS} = 100$ mV is ~ 0.5 A/W. This value was increased to ~ 5 A/W at an illumination intensity of 22 nW with $V_{GS} = 60$ V (Figure 3(d)). Variation of $R$ with respect to effective laser intensity and under different gate voltage biases ($V_{GS}$) are shown in Figure 3(d).

As mentioned earlier, layer thickness plays an important role in electrical/opto-electronic transport properties of 2D Van der Waals materials. It has been shown that mobility varies as thickness changes [20,34]. Also, the underlying substrate plays an import role in opto-electronic transport as charge traps states are primarily originate from underlying substrate [18,35]. Effect of substrate is screened in thicker channel devices [20]. Thickness dependent studies are particularly important in optoelectronic devices as absorption of light could vary in thinner samples [36]. Therefore, to investigate the thickness dependent optical properties of InSe photo-FETs, a systematic analysis as well as correlation of a variety of core intrinsic properties/performance parameters of devices with varying channel thickness were performed. Our main findings are summarized in Table 1 and Figure 4. As reported earlier [20], the field-effect mobility ($\mu_{FE}$) decreases with increasing thickness (Figure 4(a)). This behavior is due to increased interlayer coupling resistance and reduced screening in thicker FETs as seen previously in $MoS_2$ [34] as well. To determine the nature of screening from substrate charge traps, we extracted screening exponent ($\chi$) from $\sigma \propto (\Delta V_G)^\chi$, where $\Delta V_G = V_{GS} - V_{TH}$. Uncscreened charge traps corresponds to $\chi = 2$ and completely screened charge traps corresponds to $\chi = 1$ [18]. Dependence of screening exponent ($\chi$) on thickness in shown in Figure 4(b). For $t = 20$ nm, $\chi$ is 1.54 which could be due to partially screened charge traps originating because of the substrate. As thickness is increased, $\chi$ decreases which is a consequence of increased screening due to additional layers [18,20].

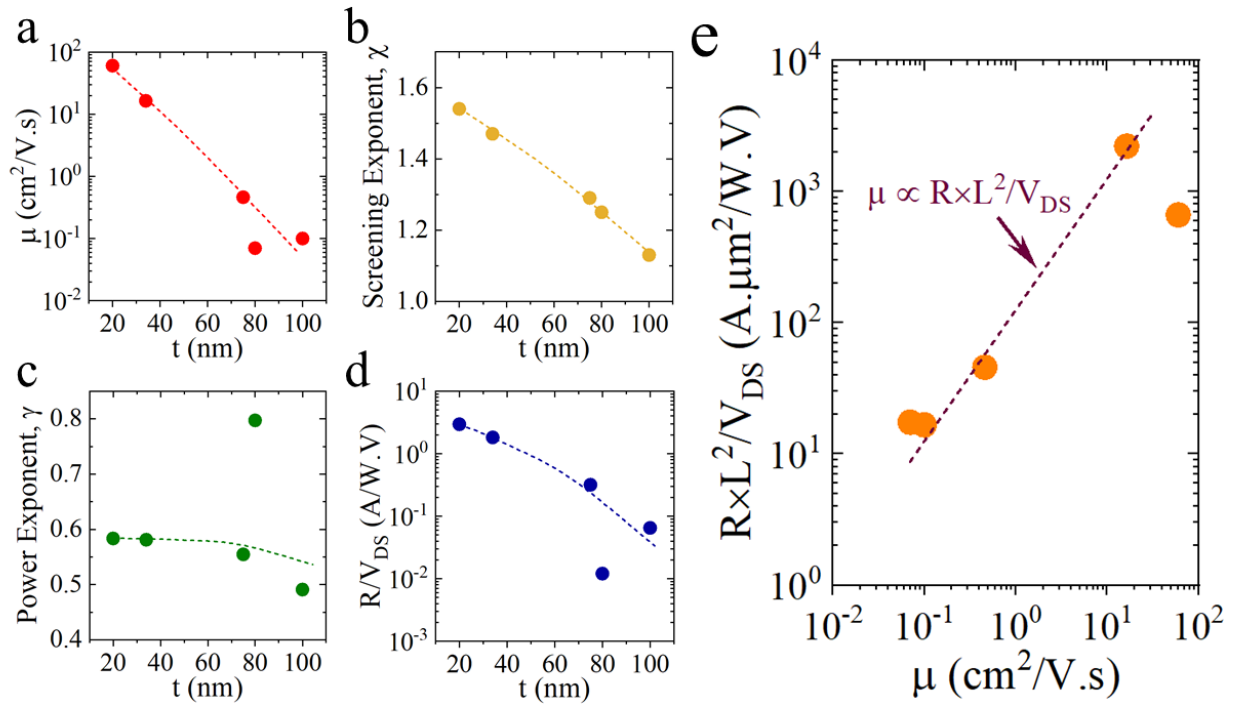

**Figure 4:** a) Variation of field-effect mobility ($\mu_{FE}$) with thickness (t), dotted line is guide to an eye. b) Variation of screening exponent ($\chi$) with thickness (t), dotted line is guide to an eye. c) Variation of power exponent ($\gamma$) with thickness (t), dotted line is guide to an eye. d) Variation of normalized responsivity ($R/V_{DS}$) with thickness (t), dotted line is guide to an eye. e) Dependence of $R \times L^2/V_{DS}$ as a function of field-effect mobility ($\mu_{FE}$). Dotted purple line correspond to $\mu_{FE} \propto R \times L^2/V_{DS}$.

Variation of power exponent ($\gamma$) on thickness is shown in Figure 4(c). These values of $\gamma$ were extracted from measurements performed without any back gate voltage. We found that for all the devices measured, $\gamma < 1$. This indicates photogating could be the dominant photocurrent generation [30] mechanism under the aforementioned measurement conditions. To have a better comparison of responsivity among the devices tested, we have normalized the responsivity ($R$) with applied drain-source bias ($V_{DS}$). Normalized responsivity ($R/V_{DS}$) decreased with increasing thickness (Figure 4(d)). It has been shown previously [37] that $R$ and $\gamma$ are inversely related. A lower $\gamma$ could correspond to stronger photogating with higher responsivity. Here we expected that $R/V_{DS}$ should increase with increasing thickness. However, we observed contrasting results. Thus, particularly when mobility of channel increases, photogating has a lesser (or secondary) effect. To assert this statement, we plotted $R \times L^2/V_{DS}$ as a function of field-effect mobility ($\mu_{FE}$) shown in Figure 4(e). Internal gain ($G_{int}$) can be determined by $G_{int} = \eta_{sep} \times \tau_{lifetime}/\tau_{transit}$, where $\eta_{sep}$ is separation efficiency, $\tau_{lifetime}$ is lifetime of minority charge traps (trapping/detrapping) and $\tau_{transit}$, is transit time of majority charge carriers [38]. Transit time depends on the mobility ($\mu_{FE}$), length of channel (L) and applied electric field ($V_{DS}/L$) as $\tau_{transit} = L^2/(V_{DS} \times \mu_{FE})$ [38]. Thus, the internal gain depends on mobility as $G_{int} = \mu_{FE} \times V_{DS} \times \eta_{sep} \times \tau_{lifetime}/L^2$. For lower illumination intensities, the external quantum efficiency (EQE) approaches internal gain ($G_{int}$) [39]. Thus, at lower effective powers, internal gain can also be determined by $G_{int} = R \times h \times c/(e \times \lambda)$ where $h$ is the planks constant, c is the speed of light, e is the charge of an electron and $\lambda$ is the wavelength of illumination light. Assuming $\eta_{sep}$ (~ 1) and $\tau_{lifetime}$ as constants, we can determine relation between field-effect mobility ($\mu_{FE}$) and responsivity ($R$) as $\mu_{FE} \propto R \times L^2/V_{DS}$ (denoted by purple dotted line Figure 4(e)). Here, we would like to point out that $\tau_{lifetime}$ could vary as thickness varies. However, if we assume $\tau_{lifetime} \propto 1/\gamma$ (as shown previously in InSe by Zhao et. al. [37]), variation in $\tau_{lifetime}$ would be minimal or of the same order as $\gamma$ is similar for devices measured. The $\tau_{lifetime}$ measured for one of our devices (D5) is ~ 20 µs as shown in supplementary data (Figure S1).

**Table 1**: InSe device parameters. $T = 300K$, $t =$ channel thickness, $L$ and $W =$ length and width of the device, $V_{DS} =$ drain-source voltage, $\mu_{FE} =$ field-effect mobility, $R =$ responsivity, $\chi =$ screening exponent, $\gamma =$ power exponent.

| Device # | $t$ (nm) | $L \times W$ (μm) | $V_{DS}$ (V) | $\mu_{FE}$ ($cm^2/Vs$) | $R/V_{DS}$ ($A/W \cdot V$) | $\chi$ | $\gamma$ |
|---|---|---|---|---|---|---|---|
| **D1** | 20 | 15 × 2.5 | 0.2 | 60 | 2.95 | 1.54 | 0.58 |
| **D2** | 34 | 35 × 60 | 0.1 | 16.34 | 1.81 | 1.47 | 0.58 |
| **D5** | 75 | 12 × 21.6 | 0.5 | 0.46 | 0.318 | 1.29 | 0.55 |
| **D6** | 80 | 38 × 43.6 | 1 | 0.07 | 0.012 | 1.25 | 0.79 |
| **D7** | 100 | 16 × 53 | 0.1 | 0.1 | 0.065 | 1.13 | 0.49 |

**Conclusion**

In conclusion, we have systematically investigated the photoconductive properties of InSe based phototransistors. The thickness dependence of the photo conducting layers and its effect on various crucial device parameters and mechanisms that could lead up to higher responsivities were discussed in detail. Our findings suggest that although the phenomenon of photogating which could lead to higher responsivities, was present in all the devices studied, other key intrinsic device properties such as field-effect mobility will play a stronger role in obtaining and/or tuning the photo responsive behavior of a particular material. We can also conclude that even if multilayer InSe (thicker) channel materials are good candidates for photon absorption, compared to other direct bandgap monolayer materials, a thorough understanding of the "ideal" thickness that will provide an optimal combination of parameters and properties required for high performance devices is needed. While our conclusions are primarily drawn from experiments performed on InSe based devices, we believe that such structure-property correlation exists in other 2D Van der Waals solids as well.

**Acknowledgements**

This work was supported by the U.S. Army Research Office MURI grant #W911NF-11-1-0362. S.T. and P.D.P. acknowledges the support from Indo-U.S. Virtual Networked Joint Center Project on "Light Induced Energy Technologies: Utilizing Promising 2D Nanomaterials (LITE UP 2D)" through the grant number IUSSTF/JC-071/2017. The scanning electron microscope used in this work was purchased through a grant from National Science Foundation (CHE 0959568). MW & P.D.P acknowledges the College of Science Dissertation Research Award and Graduate School Doctoral Fellowship respectively, awarded at Southern Illinois University Carbondale (SIUC). LW acknowledges support through SIUC's REACH and Energy Boost Awards.